\documentclass[twocolumn,showpacs,preprintnumbers,amsmath,amssymb]{revtex4}
\usepackage{dcolumn}
\usepackage{bm}
\usepackage{graphicx}
\usepackage{amsmath}
\begin{document}

\title{Coherent control of spin tunneling in a spin-orbit coupled bosonic triple well}
\author{\ Yuxin Luo$^{1}$, \ Jia Yi$^{2}$, \ Wenjuan Li$^{1}$, \ Xin Xie$^{1}$, \ Xiaobing Luo$^{3,4}$, and\ Yunrong Luo$^{1}$\footnote{Corresponding author: lyr\underline{ }1982@hunnu.edu.cn}}
\affiliation{$^{1}$Key Laboratory for Matter Microstructure and Function of Hunan Province, and Key Laboratory of
Low-dimensional Quantum Structures and Quantum Control of Ministry
of Education, School of Physics and Electronics, Hunan Normal University, Changsha 410081, China\\
$^{2}$School of Management, Hunan University of Information Technology, Changsha 410151, China\\
$^{3}$Department of Physics, Zhejiang Sci-Tech University, Hangzhou 310018, China\\
$^{4}$Department of Physics, Jinggangshan University, Ji'an 343009, China}

\begin{abstract}

We study the coherent control of spin tunneling for a spin-orbit (SO) coupled boson held in a driven triple well. Under high-frequency approximation, we analytically obtain the quasienergies of the SO-coupled bosonic triple-well system and fine energy band structure is displayed. By adjusting the driving parameters, we reveal that the directed selective spin-flipping or spin-conserving tunneling of a SO-coupled boson occurs along different pathways and in different directions. The analytical results are numerically confirmed and perfect agreements are found. Further, a scheme of quantum spin switch with or without spin-flipping is presented. These results may be useful for quantum information processing and the design of spintronic devices.

\end{abstract}

\pacs{03.65.Xp, 32.80.Qk, 68.65.Fg, 71.70.Ej}

\maketitle

\section{Introduction}

Spin-orbit (SO) coupling, the interaction between the spin and motion of a quantum particle, plays a crucial role in many exotic condensed matter phenomena and applications, such as spin-Hall effect\cite{kato306}, topological insulator\cite{bernevig314}, the persistent spin helix\cite{koralek458}, spintronic devices\cite{zutic}, etc. In cold atomic gases, SO coupling can be generated through the interaction between laser and atom, which yields Abelian or non-Abelian gauge fields for atoms in the dressed two atomic internal states. An important difference between electrons and ultracold atoms is that electrons are fermions and ultracold atoms may be bosons. Therefore, SO-coupled ultracold atomic gases will lead to novel SO physics that has not been explored in solid materials.

In recent experiments, artificial SO coupling of both bosonic and fermionic ultracold atoms has been realized\cite{lin471, wang109, cheuk109, zhang109, huang12, wu354}, which provides an ideal platform to study novel SO coupling physics, owing to their unprecedented level of controllability over all available experimental parameters. There have been a number of research works focusing on the intriguing dynamics of the SO-coupled cold atomic gases, such as Josephson dynamics of SO-coupled Bose-Einstein condensates in a double-well potential\cite{zhang609, garcia89, citro224}, selective spin transportation in a SO-coupled bosonic junction\cite{yu90}, Anderson localization of cold atomic gases with effective SO interaction in a quasiperiodic optical lattice\cite{zhou87}, Klein tunneling\cite{zhang628}, collective dynamics\cite{zhang109, chen86, hu93}, nonequilibrium dynamics of SO-coupled lattice bosons\cite{ng92}, tunable Landau-Zener transitions in SO-coupled atomic gases\cite{olson90}, spin dynamics of SO-coupled Bose-Einstein condensates in a random potential\cite{mardonov115}, controlling spin-dependent localization and directed transport in a bipartite lattice\cite{luo93}, Bloch oscillations of SO-coupled cold atoms in an optical lattice\cite{ji99}, dynamics of SO-coupled cold atomic gases in a Floquet lattice with an impurity\cite{luo52}, controlling stable tunneling in a non-Hermitian SO-coupled bosonic junction\cite{luo22}, the superflow of SO-coupled Bose-Einstein condensates in optical lattices\cite{luoxb103}, controlling directed atomic motion and second-order tunneling of a SO-coupled atom in optical lattices\cite{luo103}, and so on.

As mentioned above, many works on the tunneling dynamical properties of SO-coupled ultracold atomic gases, which have focused on systems with double-well or optical lattice potentials. To the best of our knowledge, research on the coherent control of spin tunneling for SO-coupled cold atomic gases held in a triple well is extremely rare. However, the triple-well system is a typical model to demonstrate the coherent control of spin tunneling dynamics and is an important bridge between the double well and the optical lattice systems to fully understand the spin tunneling and transport of SO-coupled cold atomic gases in the quantum wells. Thus, it motivates us to study the spin tunneling dynamics in a SO-coupled bosonic triple-well system.

In this paper, we theoretically investigate the coherent control of spin tunneling for a SO-coupled boson held in a driven triple-well potential. In high-frequency approximation, the quasienergies of the SO-coupled cold atomic triple-well system are analytically obtained and the quasienergy spectrum is shown. By adjusting the time-dependent driving field, we can manipulate the directed selective spin-flipping or non-spin-flipping tunneling of a SO-coupled boson along different pathways and in different directions. The analytical results are confirmed by direct numerical simulations and good agreements are displayed. Further, we present an interesting scheme of quantum spin tunneling switch with or without spin-flipping for transporting a SO-coupled boson from well 1 to well 3. The results may be useful in the design of spintronic devices and can be observed with the current experimental setups.

\section{Analytical solutions and quasienergy spectra in the high-frequency approximation}

We consider a single SO-coupled ultracold boson held in a driven triple-well potential, in which the dynamics of the system is governed
by a Hamiltonian \cite{luo93, li91}
\begin{eqnarray}\label{eq1}
\hat{H}(t)&=&-\nu(\hat{a}^{\dag}_{1}e^{-i\pi\gamma\hat{\sigma}_{y}}\hat{a}_{2}+\hat{a}^{\dag}_{2}e^{-i\pi\gamma\hat{\sigma}_{y}}\hat{a}_{3}+H.c.) \nonumber\\
&+&\frac{\Omega}{2}\sum_{j}(\hat{n}_{j\uparrow}-\hat{n}_{j\downarrow})+\sum_{\sigma}[\varepsilon_1(t)\hat{n}_{1\sigma}-\varepsilon_2(t)\hat{n}_{3\sigma}].\nonumber\\
\end{eqnarray}
Here $\hat{a}_{j}=(\hat{a}_{j \uparrow},\hat{a}_{j \downarrow})^{T}$ (the superscript $T$
stands for the matrix transpose) is the two-component vector with elements being the annihilation operators of spin-up and spin-down atoms in the $j$th $(j=1,2,3)$ well, and $\hat{a}_{j}^{\dag}$ denotes its Hermite conjugation with elements being the creation operators. $\nu$ denotes the tunneling amplitude without SO coupling, $\gamma$ characterizes the SO coupling strength, $\hat{\sigma}_{y}$ is the $y$ component of Pauli operator, $\Omega$ is the effective Zeeman field intensity, and $\hat{n}_{j\sigma}=\hat{a}_{j \sigma}^{\dag}\hat{a}_{j \sigma}$ denotes the number operator for spin $\sigma$ $(\sigma=\uparrow, \downarrow)$ in well $j$. The function $\varepsilon_{1,2}(t)=\varepsilon_{1,2}\cos(\omega t)$ is the periodic driving field, in which $\varepsilon_{1,2}$ and $\omega$ are the driving amplitude and frequency respectively \cite{li91}. Throughout this paper, $\hbar=1$ is adopted and the dimensionless parameters $\nu$, $\Omega$, $\varepsilon_{1,2}$, $\omega$ are in units of the reference frequency $\omega_0=0.1 E_r$, with $E_r=k_{L}^2/(2m)$ being the single-photon recoil energy, and time $t$ is normalized in units of $\omega_0^{-1}$. Note that the single-photon recoil energy is $E_r=k_{L}^2/(2m)=22.5$ kHz and the Zeeman field $\Omega$ is set as $-40 \omega_0\sim 40 \omega_0$ in the experiment\cite{lin471}, and the experimentally achievable system parameters can be tuned in a wide range as follows \cite{yu90, luo93, luo22, chen107}: $\nu \sim \omega_0$, $\varepsilon_{1,2} \sim \omega \in [0, 100](\omega_0)$, $\Omega \sim \omega$.

Using the Fock basis $|\sigma, 0, 0\rangle$ (or $|0, \sigma, 0\rangle$ or $|0, 0, \sigma\rangle$) to represent the state of a spin $\sigma$ atom occupying the well 1 (or well 2 or well 3) and no atom in the other two wells, we can expand the quantum state of the SO-coupled system as
\begin{eqnarray}\label{eq2}
|\psi(t)\rangle&=&a_1(t)|\uparrow,0,0\rangle+a_2(t)|\downarrow,0,0\rangle\nonumber\\ &+&a_3(t)|0,\uparrow,0\rangle
+a_4(t)|0,\downarrow,0\rangle\nonumber\\&+&a_5(t)|0,0,\uparrow\rangle+a_6(t)|0,0,\downarrow\rangle,
\end{eqnarray}
where $a_{k}(t)$ ($k=1, 2, ..., 6$) denotes the time-dependent probability amplitude of the boson being in the corresponding Fock state $|\sigma, 0, 0\rangle$ or $|0, \sigma, 0\rangle$ or $|0, 0, \sigma\rangle$(e.g., $a_{1}(t)$ denotes the time-dependent probability amplitude of the atom being in state $|\uparrow, 0, 0\rangle$). The corresponding probability reads $P_{k}(t)=|a_{k}(t)|^2$, which obeys the normalization condition $\sum_{k=1}^6 P_{k}(t)=1$. Inserting equations (1) and (2) into Schr\"{o}dinger equation $i\frac{\partial|\psi(t)\rangle}{\partial t}=\hat{H}(t)|\psi(t)\rangle$
results in the coupled equations
\begin{eqnarray}\label{eq3}
i\dot{a}_{1}(t)&=&[\frac{\Omega}{2}+\varepsilon_1\cos(\omega t)]a_1(t)-\nu\cos(\pi\gamma)a_3(t)\nonumber\\&+&\nu\sin(\pi\gamma)a_4(t),\nonumber\\
i\dot{a}_{2}(t)&=&[-\frac{\Omega}{2}+\varepsilon_1\cos(\omega t)]a_2(t)-\nu\sin(\pi\gamma)a_3(t)\nonumber\\&+&\nu\cos(\pi\gamma)a_4(t),\nonumber\\
i\dot{a}_{3}(t)&=&-\nu\cos(\pi\gamma)[a_1(t)+a_5(t)]-\nu\sin(\pi\gamma)\nonumber\\ & \times &[a_2(t)-a_6(t)]+\frac{\Omega}{2}a_3(t),\nonumber\\
i\dot{a}_{4}(t)&=&-\nu\sin(\pi\gamma)[a_1(t)+a_5(t)]-\nu\cos(\pi\gamma)\nonumber\\ & \times &[a_2(t)+a_6(t)]-\frac{\Omega}{2}a_4(t),\nonumber\\
i\dot{a}_{5}(t)&=&[\frac{\Omega}{2}-\varepsilon_2\cos(\omega t)]a_5(t)-\nu\cos(\pi\gamma)a_3(t)\nonumber\\&-&\nu\sin(\pi\gamma)a_4(t),\nonumber\\
i\dot{a}_{6}(t)&=&[-\frac{\Omega}{2}-\varepsilon_2\cos(\omega t)]a_6(t)+\nu\sin(\pi\gamma)a_3(t)\nonumber\\&-&\nu\cos(\pi\gamma)a_4(t).
\end{eqnarray}

It is difficult to obtain the exact analytical solutions of equation (3), because of the periodically varying coefficients. However, in the high-frequency approximation\cite{blanes470,thimmel9}, it can become a set of linear equations with constant coefficients, which is analytically solvable. Therefore, we introduce the slowly varying function of time $b_{k}(t) (k=1,2,...,6)$ through the transformation
$a_1(t)=b_1(t)e^{-i\int[\frac{\Omega}{2}+\varepsilon_1\cos(\omega t)]dt}$,
$a_2(t)=b_2(t)e^{-i\int[-\frac{\Omega}{2}+\varepsilon_1\cos(\omega t)]dt}$,
$a_3(t)=b_3(t)e^{-i\int\frac{\Omega}{2}dt}$,
$a_4(t)=b_4(t)e^{i\int\frac{\Omega}{2}dt}$,
$a_5(t)=b_5(t)e^{-i\int[\frac{\Omega}{2}-\varepsilon_2\cos(\omega t)]dt}$,
$a_6(t)=b_6(t)e^{i\int[\frac{\Omega}{2}+\varepsilon_2\cos(\omega t)]dt}$. Under the high-frequency approximation and by using of the Fourier expansion $ e^{\pm i\int \varepsilon_{1,2}(t)dt}=\sum_{n=-\infty}^{\infty}\mathcal{J}_{n}(\frac{\varepsilon_{1,2}}{\omega})e^{\pm i n \omega t}$ and $e^{\pm i\int[\varepsilon_{1,2}(t)\pm \Omega]dt}=\sum_{n'=-\infty}^{\infty}\mathcal{J}_{n'}(\frac{\varepsilon_{1,2}}{\omega})e^{\pm i(n'\pm \frac{\Omega}{\omega})\omega t}$, we can neglect these rapidly oscillating terms of the Fourier expansion with $n\neq 0$ and $n'\pm \frac{\Omega}{\omega}\neq 0$\cite{zou46} and the set of differential equation (3) is transformed to the form
\begin{eqnarray}\label{eq4}
i\dot{b}_{1}(t)&=&-J_{1}b_3(t)+J_{2} b_4(t),\nonumber\\
i\dot{b}_{2}(t)&=&-J_{3}b_3(t)-J_{1} b_4(t),\nonumber\\
i\dot{b}_{3}(t)&=&-J_{1}b_1(t)-J_{3} b_2(t)-J_{4} b_5(t)+J_{5} b_6(t),\nonumber\\
i\dot{b}_{4}(t)&=&J_{2}b_1(t)-J_{1} b_2(t)-J_{6} b_5(t)-J_{4} b_6(t),\nonumber\\
i\dot{b}_{5}(t)&=&-J_{4}b_3(t)-J_{6} b_4(t),\nonumber\\
i\dot{b}_{6}(t)&=&J_{5}b_3(t)-J_{4} b_4(t),
\end{eqnarray}
where the effective coupling constants are written as $J_{1}=\nu \cos(\pi \gamma)\mathcal{J}_{0}(\frac{\varepsilon_1}{\omega})$,
$J_{2,3}=\nu \sin(\pi \gamma)\mathcal{J}_{\mp\frac{\Omega}{\omega}}(\frac{\varepsilon_1}{\omega})$,
$J_{4}=\nu \cos(\pi \gamma)\mathcal{J}_{0}(\frac{\varepsilon_2}{\omega})$, and
$J_{5,6}=\nu \sin(\pi \gamma)\mathcal{J}_{\mp\frac{\Omega}{\omega}}(\frac{\varepsilon_2}{\omega})$
with $\mathcal{J}_{n}(x)$ being the $n$-order Bessel function of $x$. Equation (4) is effective in the description of the spin tunneling
dynamics of the original system, which is the basis of the following analysis.

\begin{figure}
\includegraphics[height=1.3in,width=1.6in]{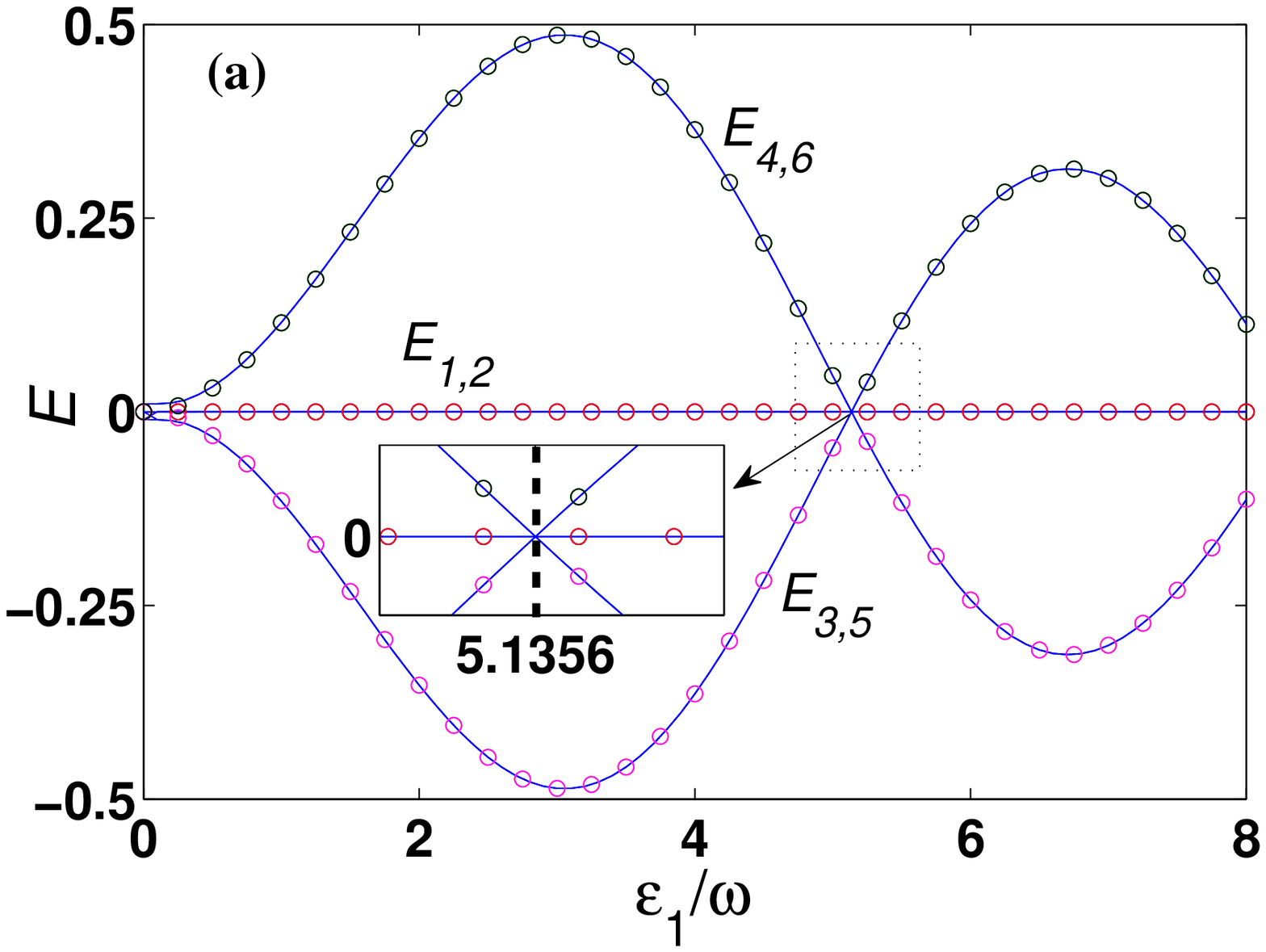}
\includegraphics[height=1.3in,width=1.6in]{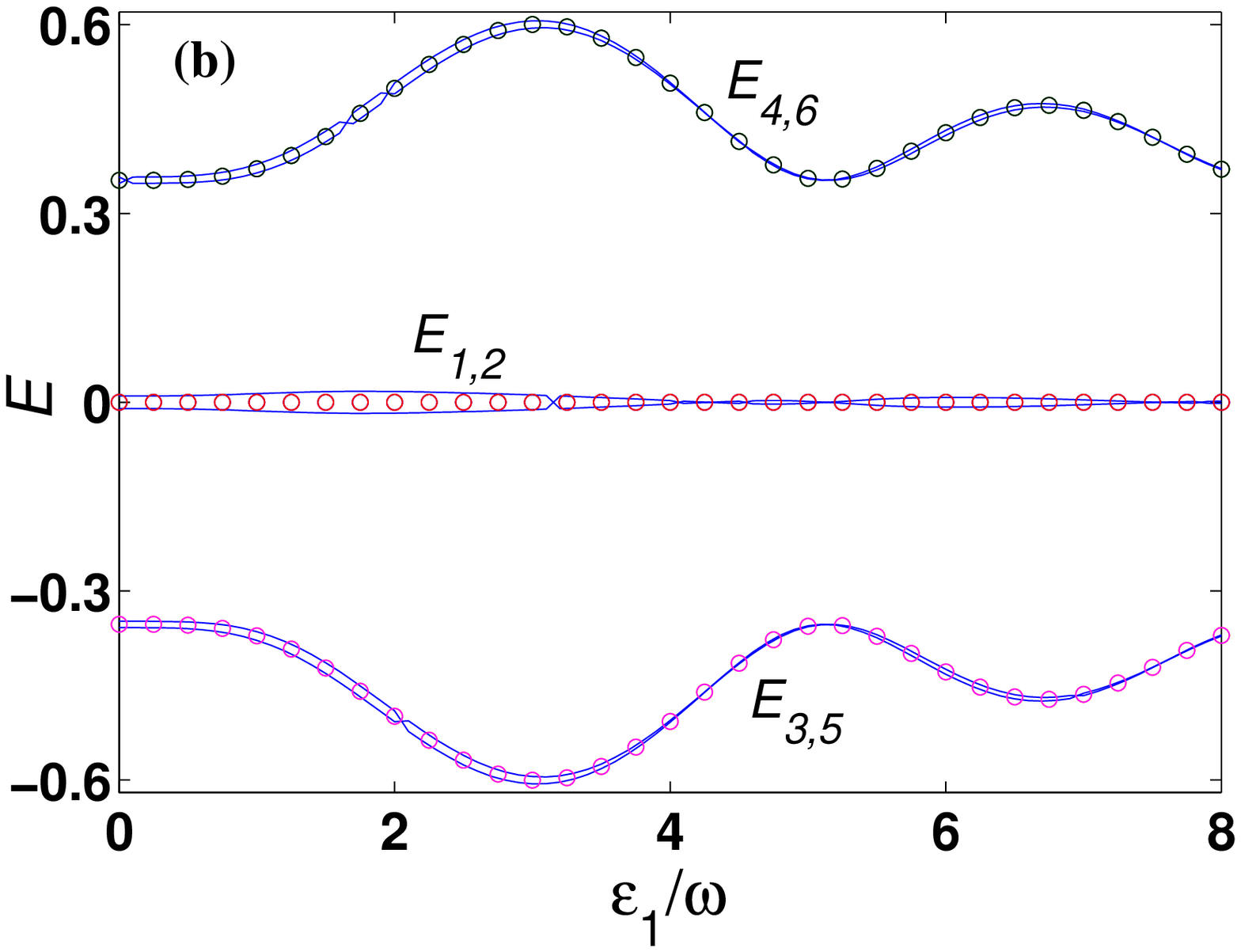}
\caption{\scriptsize{(Color online) Quasienergy as a function of the driving parameter $\varepsilon_1/\omega$ for $\gamma=0.5, \nu=1, \omega=50, \Omega=100$, and (a) $\varepsilon_2=5.1356 \omega$; (b) $\varepsilon_2=2 \omega$. Here, circles label the analytical results and solid curves denote the numerical correspondences. Hereafter, any parameter adopted in the figures is dimensionless.}}
\end{figure}

Based on the Floquet theorem\cite{shirley138}, the analytical Floquet state, non-Floquet state, and quasienergies can be obtained from equation (4).
The Floquet solution is constructed as $|\psi(t)\rangle=|\varphi(t)\rangle e^{-i E t}$, where $E$ is the Floquet quasienergy and
$|\varphi(t)\rangle=A_1 e^{-i\int[\frac{\Omega}{2}+\varepsilon_1\cos(\omega t)]dt}|\uparrow,0,0\rangle+A_2 e^{-i\int[-\frac{\Omega}{2}+\varepsilon_1\cos(\omega t)]dt}|\downarrow,0,0\rangle + A_3 e^{-i\int\frac{\Omega}{2}dt}|0,\uparrow,0\rangle
+A_4 e^{i\int\frac{\Omega}{2}dt}|0,\downarrow,0\rangle+A_5 e^{-i\int[\frac{\Omega}{2}-\varepsilon_2\cos(\omega t)]dt}|0,0,\uparrow\rangle+A_6 e^{i\int[\frac{\Omega}{2}+\varepsilon_2\cos(\omega t)]dt}|0,0,\downarrow\rangle$ is the Floquet state with the same period as the Hamiltonian (1). According to the superposition principle of quantum mechanics, the non-Floquet state can be constructed by the linear superposition of the Floquet states, which implies a quantum interference effect among the Floquet states with different quasienergy\cite{luo22, lu379, luo48, luo7}. We have introduced the stationary solution of equation (4) as $b_{k}(t)=A_{k} e^{-i E t}$ with $A_{k}$ being constant. Inserting such a form of $b_{k}(t)$ into equation (4), the constant $A_{k}$ and Floquet quasienergy $E$ can be obtained. Here, we give the Floquet quasienergies as
\begin{eqnarray}\label{eq5}
E_{1,2}=0, E_{3,4}=\mp\frac{\sqrt{\alpha-\sqrt{\beta}}}{\sqrt{2}},
E_{5,6}=\mp\frac{\sqrt{\alpha+\sqrt{\beta}}}{\sqrt{2}},
\end{eqnarray}
with the constants $\alpha=2J_{1}^{2}+J_{2}^{2}+J_{3}^{2}+2J_{4}^{2}+J_{5}^{2}+J_{6}^{2}$ and
$\beta=J_{2}^{4}+4J_{1}^{2}(J_{2}-J_{3})^{2}+J_{3}^{4}+2J_{3}^{2}J_{5}^{2}+4J_{4}^{2}J_{5}^{2}+J_{5}^{4}+8J_{1}J_{4}(J_{2}-J_{3})(J_{5}-J_{6})-8J_{4}^{2}J_{5}J_{6}+J_{6}^{4}
-2J_{6}^{2}(J_{3}^{2}-2J_{4}^{2}+J_{5}^{2})-2J_{2}^{2}(J_{3}^{2}+J_{5}^{2}-J_{6}^{2})$ being adjusted by the system parameters, so that all the quasienergies are determined for a set of fixed parameters except for the degenerate zero quasienergies $E_1$ and $E_2$.

From equation (5), it is surprisingly found that when the SO coupling strength $\gamma$ is integer or half integer, the constant $\beta$ will equal to zero. This will result in the quasienergies $E_3=E_5=-\sqrt{\alpha/2}$ and $E_4=E_6=\sqrt{\alpha/2}$, which means that the new quasienergy degeneracy occurs. To our knowledge, the degeneracy of energy levels generally implies quantum decoherence, so it will lead to the intriguing phenomenon of quantum tunneling, for instance, the selective coherent destruction of tunneling (SCDT) takes place at the degeneracy (crossing) point of the partial energy levels \cite{lu379, luo48, luo7} and coherent destruction of tunneling (CDT) occurs at the collapse (crossing) point of all energy levels\cite{grossmann, liu377}.
As an example, we set the parameters $\gamma=0.5, \omega=50, \Omega=100, \nu=1$, and (a) $\varepsilon_2=5.1356 \omega$, (b) $\varepsilon_2=2 \omega$ to plot the quasienergy spectra with quasienergy as a function of the driving parameter $\varepsilon_1/\omega$, as in figure 1(a)-(b), respectively.
Here, the circle points denote the analytical results based on equation (5) and the curves label the numerical results from the exact model (3). The two sets of parameters lead to $E_3=E_5=-|\mathcal{J}_{2}(\frac{\varepsilon_1}{\omega})|$ and $E_4=E_6=|\mathcal{J}_{2}(\frac{\varepsilon_1}{\omega})|$, and $E_3=E_5=-\sqrt{\mathcal{J}_{2}(\frac{\varepsilon_1}{\omega})^2+\mathcal{J}_{2}(2)^2}$ and $E_4=E_6=\sqrt{\mathcal{J}_{2}(\frac{\varepsilon_1}{\omega})^2+\mathcal{J}_{2}(2)^2}$ in equation (5), respectively. In figure 1 (a), we observe that the curves of quasienergies $E_1$ and $E_2$, $E_3$ and $E_5$, $E_4$ and $E_6$ are degenerate respectively. The numerical results (curves) are in good agreement with the analytical ones (circles). Not only that, there exists some collapse (crossing) points of the quasienergy spectra (i.e., see the amplified inset of figure 1(a)), which correspond to the roots of equation $\mathcal{J}_{2}(\varepsilon_1/\omega)=0$ (i.e., $\mathcal{J}_{2}(\varepsilon_1/\omega)=\mathcal{J}_{2}(5.1356)=0$). Compared to the single energy band structure shown in figure 1 (a), the quasienergy curves are divided to three energy bands in figure 1 (b). The energy gap between two adjacent energy bands is equal to the minimum value $\mathcal{J}_{2}(2)\approx 0.3528$ of quasienergy $E_{4}$ or $E_{6}$. Furthermore, from figure 1 (b) it can be seen that
the two energy-level curves of each band from the numerical results are approximately degenerate (e.g., quasienergy curves of $E_1$ and $E_2$, $E_3$ and $E_5$, $E_4$ and $E_6$), which have some small deviations from analytical results (circles).

\section{Directed tunneling and spin switch with or without spin-flipping}

Coherent control of quantum tunneling is an interesting and important research subject and possesses many potential applications in quantum information technologies\cite{monroe, kral}. Based on the fact that the spin tunneling dynamics of a SO-coupled cold atom depends strongly on the periodic driving external field in a triple well, the quantum spin tunneling and transport of a SO-coupled atom can be manipulated via adjusting the driving parameters. In the section, we will focus on studying the directed tunneling and spin switch with or without spin-flipping.

\subsection{Directed tunneling with or without spin-flipping}

\begin{figure*}[htp]\centering
\includegraphics[height=1.3in,width=2.2in]{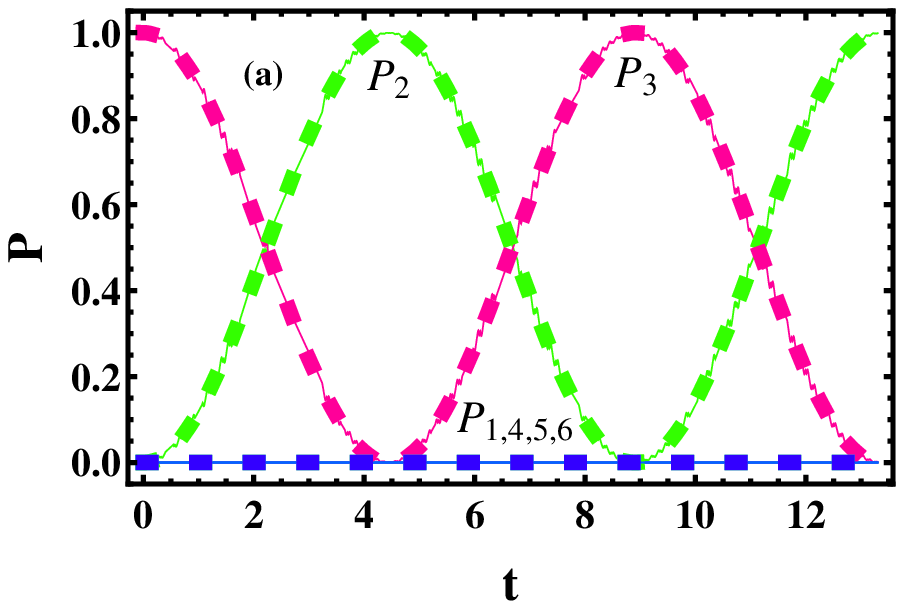}
\includegraphics[height=1.3in,width=2.2in]{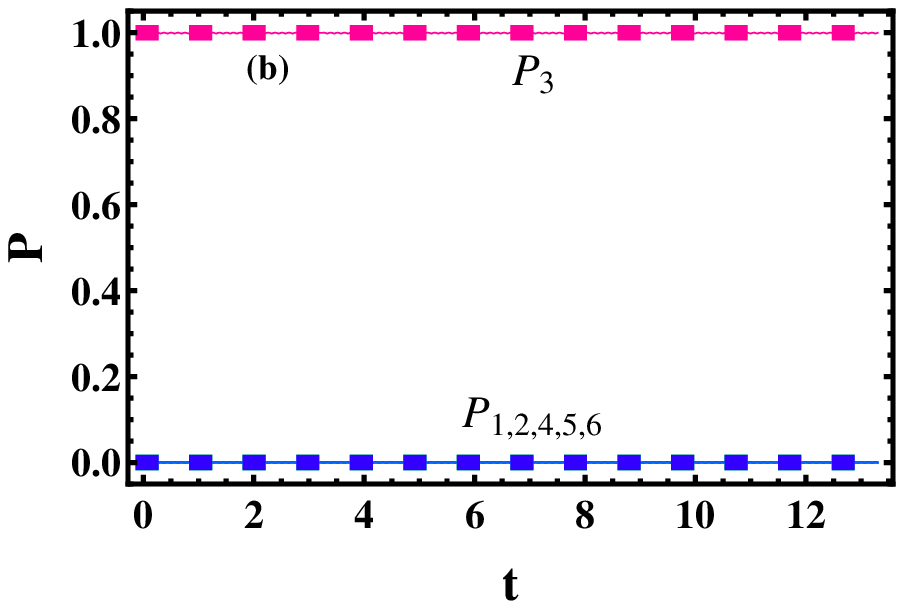}
\includegraphics[height=1.3in,width=2.2in]{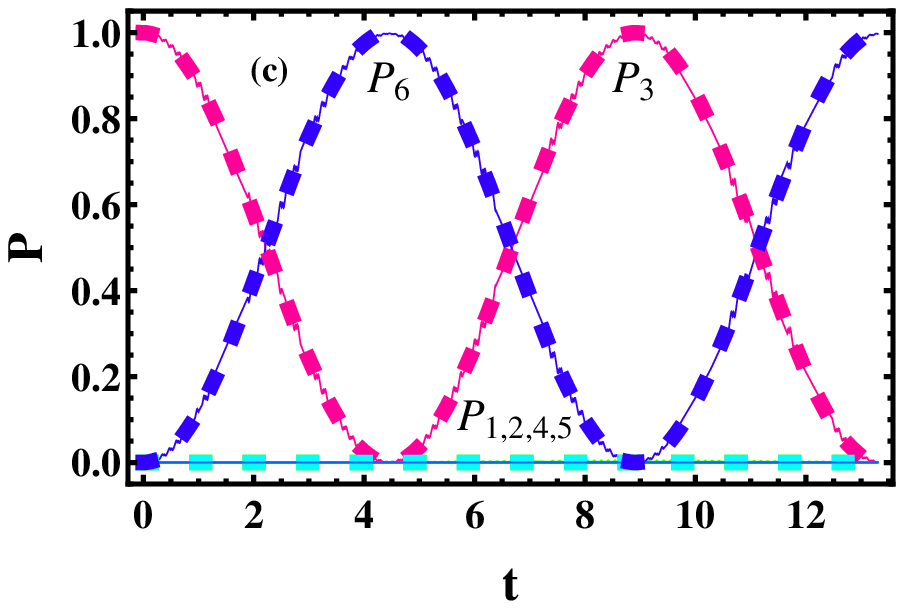}
\caption{\scriptsize{(Color online) Time evolutions of the probabilities $P_{k}(t)$ for the system parameters $\gamma=0.5, \nu=1, \omega=50, \Omega=100$, and (a) $\varepsilon_1=2 \omega$, $\varepsilon_2=5.1356 \omega$; (b) $\varepsilon_1=\varepsilon_2=5.1356 \omega$; (c) $\varepsilon_1=5.1356 \omega$, $\varepsilon_2=2 \omega$, starting the system with a spin-up particle in the well 2. Dashed lines label the analytical results and solid curves denote the numerical correspondences.}}
\end{figure*}

\begin{figure*}
\includegraphics[height=1.3in,width=2.2in]{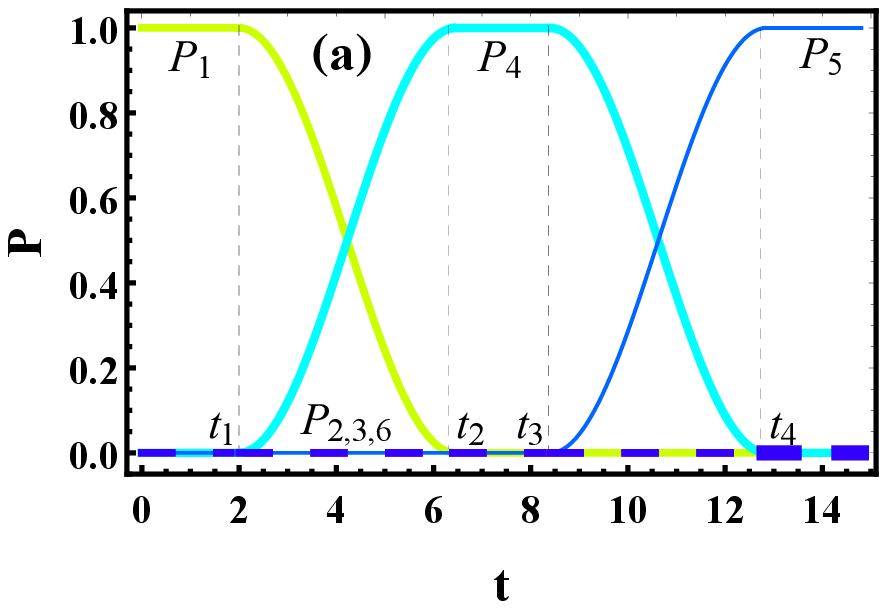}
\includegraphics[height=1.3in,width=4.2in]{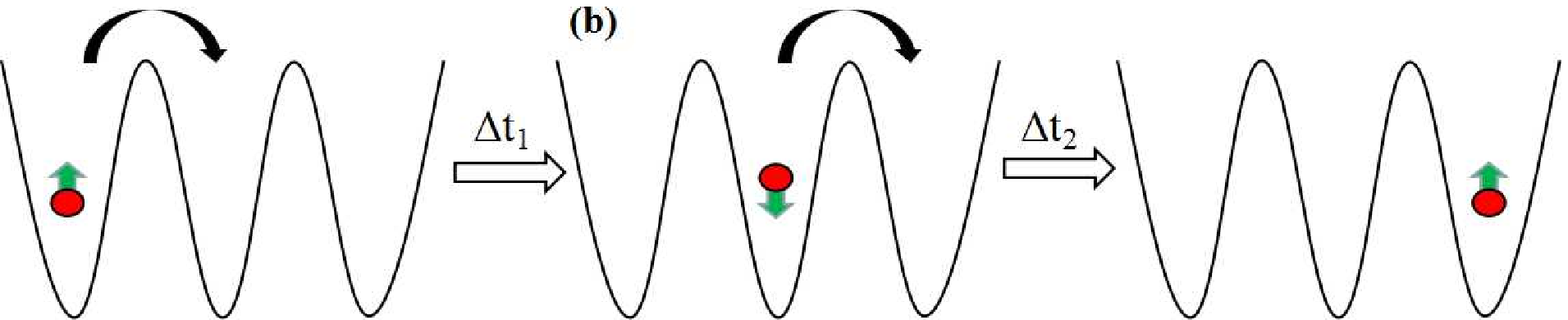}
\caption{\scriptsize{(Color online) (a) Time evolutions of the probabilities $P_{k}(t)$ for the system parameters $\gamma=0.5, \nu=1, \omega=50, \Omega=100$, and $\varepsilon_1=\varepsilon_2=5.1356\omega$ in the time intervals $0\leq t <t_1=2$, $t_2=6.4\leq t <t_3=8.4$, $t \geq t_4=12.8$, and $\varepsilon_1=2 \omega$, $\varepsilon_2=5.1356\omega$ in the time interval $t_1=2\leq t <t_2=6.4$, and $\varepsilon_1=5.1356\omega$, $\varepsilon_2=2 \omega$ in the time interval $t_3=8.4\leq t <t_4=12.8$, starting the system with a spin-up particle in the well 1. (b) Schematic diagram of the scheme for spin-flipping transporting a spin-up particle from well 1 to well 3. The time intervals $\triangle t_1=\triangle t_2=t_2-t_1=t_4-t_3=4.4$ denote the transferring times between the different populations.}}
\end{figure*}

For a SO-coupled cold atomic triple-well system, an attractive physical problem is how to control the directed selective spin-flipping or spin-conserving tunneling of a SO-coupled atom from initial middle well to left well or to right well. Based on equation (4), it can be seen that when the SO-coupling strength $\gamma$ is half integer or integer, quantum tunneling with or without spin-flipping can occur. Here, we will try to manipulate the directed selective spin-flipping tunneling of a spin-up boson initially located in the well 2 as an example. In order to realize the directed spin-flipping tunneling of the spin-up particle from well 2 to well 1, it is easily found from equation (4) that when the effective coupling constants satisfy $J_1=J_4=J_5=0$ and $J_3\neq 0$, $b_3$ is only coupled with $b_2$. It means that the SCDT occurs. We take the parameters $\gamma=0.5, \nu=1, \omega=50, \Omega=100, \varepsilon_1=2 \omega, \varepsilon_2=5.1356 \omega$ satisfying the conditions $J_1=J_4=J_5=0$ and $J_3\neq 0$ to plot the time evolutions of the probabilities in figure 2 (a). Clearly, the spin-flipping tunneling passage between well 2 and well 3 is shut off and the particle performs a spin-flipping Rabi oscillation along this pathway between well 2 and well 1, namely, the directed selective spin-flipping tunneling takes place between initial state $|0,\uparrow,0\rangle$ and state $|\downarrow,0,0\rangle$ with tunneling time $\triangle t \approx 4.4$. It means the occurrence of SCDT and the set of parameters corresponds to the degeneracy position of the partial quasienergies in figure 1 (a). If we set the same parameters as that in figure 2 (a) except for $\varepsilon_1=5.1356 \omega$, which correspond to the collapse (crossing) point of all the quasienergies in figure 1 (a), the quantum tunneling of the particle will be frozen. It means the CDT occurs as shown in figure 2 (b).

Further, from equation (4) it can also be found when the effective coupling constants satisfy $J_1=J_3=J_4=0$ and $J_5\neq 0$, the probability function $b_3$ is only related to function $b_6$, which means the SCDT happens. We fix the same initial condition and parameters as that in figure 2 (a) except for $\varepsilon_1=5.1356 \omega$ and $\varepsilon_2=2 \omega$ to plot the time evolutions of the probabilities in figure 2 (c). It is clearly seen that the directed selective spin-flipping tunneling along another pathway between well 2 and well 3 occurs, in which the spin particle performs a Rabi oscillation with spin-flipping between state $|0,\uparrow,0\rangle$ and state $|0,0,\downarrow\rangle$ with tunneling time $\triangle t \approx 4.4$. It means the SCDT occurs and the set of parameters corresponds to the degeneracy location of the partial quasienergies in figure 1 (b). Here, we note that the analytical results (dashed lines) based on equation (4) are in good agreement with the numerical ones from equation (1) in figure 2. Because the above-mentioned results are related to control the directed spin-flipping tunneling of a spin boson, it is similar to the case of manipulating the directed spin-conserving tunneling of the particle.

\subsection{spin switch with or without spin-flipping}

From the above subsection, we find the time-dependent driving field affects dramatically spin tunneling dynamics of this system, thus we can design the quantum spin switch with or without spin-flipping by applying a sequence of sudden modulation of driving strength to control opening or closing of quantum tunneling. The scheme of sudden modulation of driving parameters has been performed in many research works\cite{weiss, creffield, hai82, luo84}. Here, we present a scheme of quantum spin switch with spin-flipping by adjusting driving strength as an example, as shown in figure 3. In figure 3 (a), let a spin-up boson occupy the state $|\uparrow,0,0\rangle$ initially and fix the parameters $\gamma=0.5, \nu=1, \omega=50, \Omega=100, \varepsilon_1=\varepsilon_2=5.1356\omega$ corresponding to the collapse (crossing) point of all the quasienergies in figure 1 (a). One can see that the initially occupied state is kept, due to the CDT effect. At an arbitrarily given time $t=t_1=2$, we change the driving strength $\varepsilon_1$ to $\varepsilon_1=2 \omega$ and keep this value until $t=t_2=6.4$ at which the spin-up particle transits completely from state $|\uparrow,0,0\rangle$ to state $|0,\downarrow,0\rangle$. Then we return the driving strength $\varepsilon_1$ to the initial value such that the state $|0,\downarrow,0\rangle$ is frozen which is attributed to the effect of CDT. At an arbitrarily given time $t=t_3=8.4$, we change the driving strength $\varepsilon_2$ to $\varepsilon_2=2 \omega$ and keep this value until $t=t_4=12.8$. At this time, the spin-down particle transits completely from state $|0,\downarrow,0\rangle$ to state $|0,0,\uparrow\rangle$. Then, we return the driving strength $\varepsilon_2$ to $\varepsilon_2=5.1356 \omega$, such that the CDT effect makes the final state $|0,0,\uparrow\rangle$ is kept. So the spin-up boson is successfully transported through two spin-flipping tunnels from well 1 to well 3 by adjusting the driving strength $\varepsilon_i$ ($i=1,2$) of the time-dependent driving fields, namely, the quantum spin tunneling switch with spin-flipping is theoretically realized. The spatial distributions of the spin particle at the tuning moments are exhibited in figure 3 (b), where $\triangle t_i$ denotes transferring time between the different populations and the transferring times $\triangle t_1=\triangle t_2=t_2-t_1=t_4-t_3=4.4$. Similarly, the quantum spin tunneling switch without spin-flipping can also be performed by modulating the driving fields.

\section{conclusion and discussion}

In summary, we have studied the coherent control of spin tunneling for a SO-coupled boson trapped in a driven triple well. In the high-frequency regime, we analytically obtain the Floquet quasienergies of the system and the fine quasienergy spectrum is shown. By adjusting the strength of driving external field, we can control the directed selective spin-flipping or non-spin-flipping tunneling of a SO-coupled boson along different pathways and in different directions from the middle well to one of the other two wells. The analytical results are demonstrated by direct numerical simulations and good agreements are shown. Further, based on the combined effect of CDT and SCDT, we propose an interesting scheme of quantum spin switch that a spin-up particle is transported with or without spin-flipping from the initial well 1 to the final well 3. These results may be useful for quantum information processing and the design of quantum spintronic devices. We expect that the future experiments along the lines sketched here can be considered, which may help us further to gain new insight into the quantum spin tunneling dynamics of SO-coupled ultracold gases in multi-well systems.

\section*{ACKNOWLEDGMENTS}
This work was supported by the Hunan Provincial Natural Science Foundation of China under Grants No. 2021JJ30435 and No. 2017JJ3208, the National Natural Science Foundation of China under Grant No. 11747034, and the Scientific Research Fund of Hunan Provincial Education Department under Grant No. 18C0027. Xiaobing Luo was supported by the Scientific and Technological Research Fund of Jiangxi Provincial Education Department (numbers GJJ180559), and Open Research Fund Program of the State Key Laboratory of Low-Dimensional Quantum Physics (KF201903).

\end{document}